\documentclass[pra,twocolumn,showpacs,superscriptaddress,amsfonts,amsmath,floatfix,aps,10pt]{revtex4-1}

\usepackage{graphicx,color}

\newcommand{\sgn}{\mathop{\rm sgn}\nolimits}

\begin{document}
\title{Exact ground state properties of the one-dimensional Coulomb gas}
\author{G.E. Astrakharchik}
\affiliation{Departament de F\'{\i}sica i Enginyeria Nuclear, Campus Nord B4, Universitat Polit\`ecnica de Catalunya, E-08034 Barcelona, Spain}
\author{M. D. Girardeau}
\affiliation{College of Optical Sciences, University of Arizona, Tucson, AZ 85721, USA}
\date{\today}
\begin{abstract}
The ground state properties of a single-component one-dimensional Coulomb gas are investigated. We use Bose-Fermi mapping for the ground state wave function which permits to solve the Fermi sign problem in the following respects (i) the nodal surface is known, permitting exact calculations (ii) evaluation of determinants is avoided, reducing the numerical complexity to that of a bosonic system, thus allowing simulation of a large number of fermions. Due to the mapping the energy and local properties in one-dimensional Coulomb systems are exactly the same for Bose-Einstein and Fermi-Dirac statistics. The exact ground state energy has been calculated in homogeneous and trapped geometries by using the diffusion Monte Carlo method. We show that in the low-density Wigner crystal limit an elementary low-lying excitation is a plasmon, which is to be contrasted with the large-density ideal Fermi gas/Tonks-Girardeau limit, where low lying excitations are phonons. Exact density profiles are confronted to the ones calculated within the local density approximation which predicts a change from a semicircular to inverted parabolic shape of the density profile as the value of the charge is increased.
\end{abstract}
\maketitle


The recent progress in nanoscale technology made it possible to realize clean one-dimensional quantum gases, such as ultracold atoms confined in elongated traps\cite{exp:atomic_gases}, electrons in single-well carbon nanotubes\cite{exp:nanotubes} and in semiconductor quantum wires\cite{exp:quantum_wires}. A peculiarity of a one-dimensional world is that such systems can not be explained by the conventional Landau theory of normal Fermi liquids, but rather by an effective low-energy Luttinger liquid description\cite{Haldane}, generalizable to long-range Coulomb interactions\cite{Schulz}. A number of approaches of increasing accuracy (random-phase approximation, Singwi-Tosi-Land-Sj\"olander scheme, density functional theory) have been devised to study energetic and structural properties\cite{STLS}. The most precise calculations of energy have been obtained by the Monte Carlo technique as in Ref.~\cite{MC} where a quasi-one-dimensional geometry with a finite width of the transverse confinement is studied. The purpose of the present work is to carry out exact calculations in a strictly one-dimensional geometry.

In this paper we use the Bose-Fermi mapping to find the ground state energy of a one-dimensional single component Coulomb system exactly within statistical precision. The mapping applies both to homogeneous and trapped geometries.


We consider a single-component system of $N$ particles (bosons or fermions) of charge $e$ and mass $m$ in a one-dimensional box of length $L$. Periodic boundary conditions are applied and the Coulomb potential is truncated for interpartcle distances larger than $L/2$. The Hamiltonian reads as
\begin{equation}
\hat{H}=-\frac{\hbar^2}{2m}\sum_{i=1}^N\frac{\partial^2}{\partial z_i^2}
+\sum_{i<j}^N\frac{e^2}{|z_i-z_j|}\ .
\label{H}
\end{equation}
Natural length scales in a homogeneous system are defined by atomic units, that is Bohr radius $a_0=\hbar^2/me^2$ for length and the Rydberg $Ry =e^2/2a_0$ for energy. The system properties are governed by a single parameter, fixed by the ratio of Bohr $a_0$ and Wigner-Seitz $r_s=1/2n$ radiuses or, equivalently, by the dimensionless density $na_0$ with $n=N/L$.

Hamiltonian (\ref{H}) describes a system with the pure Coulomb interaction $V_{int}(r)=e^2/r$ between charges not screened while the movement of charges confined to a quantum wire, so that excitations of the levels of the transverse direction is not possible and the system behaves as quasi-one dimensional. For a harmonic transverse confinement this implies that the energy associated with the Hamiltonian~(\ref{H}) is small compared to $N\hbar^2/mb^2$, where $b$ is the transverse oscillator length. The peculiarity of the Coulomb interaction in a one-dimensional geometry is that its Fourier transform cannot be calculated due to a divergence at short distances. Commonly, this ultraviolet divergence is cured by considering the ground state of the transverse harmonic oscillator. This effectively introduces a cutoff to the interaction potential and removes the divergence as $V^{eff}_{int}(0)=\sqrt{\pi}e^2/b$. Here, instead we note that the unbound potential in (\ref{H}) will not produce any divergence in the energy while the pair distribution function vanishes when two particles approach each other. This permits us to successfully apply the one-dimensional Bose-Fermi mapping \cite{Girardeau60} and relate the many-body ground-state wave functions of fermions $\psi_F(z_1,\cdots,z_N)$ and bosons $\psi_B(z_1,\cdots,z_N)$ as
\begin{eqnarray}
\psi_B(z_1,\cdots,z_N) = |\psi_F(z_1,\cdots,z_N)|\ .
\label{mapping}
\end{eqnarray}
In the case of bosons the divergent $1/r$ interaction mimics the Pauli exclusion principle, which for fermions prohibits any two fermions from overlapping.


We resort to the diffusion Monte Carlo (DMC) technique for finding the ground state properties of the system. The guiding wave function is chosen in bosonic Bijl-Jastrow form, adding antisymmetrization in the case of fermions:
\begin{eqnarray}
\psi_F(z_1,\cdots,z_N) =
\prod_{i<j}^Nf_2(|z_i-z_j|)\sgn(z_i-z_j)\ .
\label{wf}
\end{eqnarray}
It is well-known that simulations of fermionic systems suffer from the ``sign problem''. In terms of the DMC algorithm this means that a ``fixed-node'' approximation should be used, that is the fermionic ground state wave function is expressed as some non-negative function multiplied by a function which has a certain nodal surface. The DMC method finds the best energy for a given choice of the nodal surface, providing an upper bound for the ground state energy. The exact ground state energy is reached only if the nodal surface is exact. Importantly, the nodal surface of Eq.~(\ref{wf}) is exact. This means that the mapping~(\ref{wf}) permits to find the {\it exact} energy of the fermionic system in DMC calculation. Furthermore, the same mapping allows avoiding evaluation of determinants and reduces the complexity of the calculation from $N^3$ to $N^2$, making accessible a larger number of particles.

We choose the two-body correlation term $f_2(z)$ as
\begin{equation}
f_2(z)=\left\{
\begin{array}{cc}
C_1\sqrt{z}I_{1}(2\sqrt{z}), & |z|<R_{par} \\
\sin^{C_2}(\pi |z|/L), & |z|>R_{par}
\end{array}
\right.\ ,
\label{f2}
\end{equation}
where coefficients $C_1,C_2$ are fixed by the continuity conditions for $f_2(z)$ and $f'_2(z)$ at the matching distance $R_{par}$. The short-range part of $f_2(z)$ corresponds to the zero-energy solution of the two-body scattering problem. This avoids divergences in the energy as two particles come close to each other. The long-range part of $f_2(z)$ is taken from a hydrodynamic theory \cite{Reatto}. The parameter $R_{par}$ is optimized by minimization of a variational energy.


The typical kinetic energy per particle scales quadratically with the density, opposite to the linear density dependence in the Coulomb interaction energy. At large densities the kinetic energy becomes dominant and can be approximated by the energy of an ideal Fermi gas
\begin{equation}
\frac{E^{IFG/TG}}{N} = \frac{\pi^2\hbar^2n^2}{6m}
\label{ETG}\ .
\end{equation}


In the opposite regime of low density the energy can be obtained by summation of the Coulomb potential energy under the assumption that all particles are localized at lattice sites of a Wigner crystal. The leading contribution to the energy (see, for example, Ref.~\cite{Dubin})
\begin{equation}
\frac{E^{Wigner}}{N}=e^2n \ln N
\label{Ewigner}
\end{equation}
for a fixed number of particles is linear in the density. Instead, for a fixed density the energy per particle and the chemical potential diverge logarithmically as the number of particles increases. This is a consequence of the long-range nature of the Coulomb interaction potential. In a three-dimensional system a similar divergence can be cured by the ``jellium'' model, that is, by restoring charge neutrality by adding a uniform charge of the opposite sign. In a one-dimensional system introduction of a uniform opposite charge would lead to a logarithmic divergence in the short-range part of the Coulomb interaction potential.

\begin{figure}
\begin{center}
\includegraphics[width=0.52\columnwidth, angle=-90]{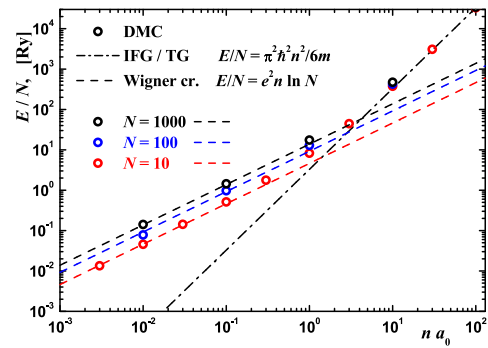}
\caption{(Color online) Energy per particle as a function of the dimensionless density for $N= 1000; 100; 10$ (from top to bottom) particles. Circles, Monte Carlo results; dash-dotted line IFG/TG energy, Eq.~(\ref{ETG}); dashed lines, energy of a Wigner crystal, Eq.~(\ref{Ewigner}).}
\label{Fig:E}
\end{center}
\end{figure}

The energy of a homogeneous system is shown in Fig.~\ref{Fig:E}. At low densities the energetic properties are precisely captured by the Wigner crystal description. At the same time no true crystal is formed as quantum fluctuations destroy the diagonal long-range order in a one-dimensional geometry\cite{Mermin}. The large-density regime is well described by an ideal Fermi gas (IFG) or Tonks-Girardeau (TG) gas depending on particle statistics. There is a smooth crossover between the Wigner crystal and IFG/TG gas regimes.

\begin{figure}
\begin{center}
\includegraphics[width=0.52\columnwidth, angle=-90]{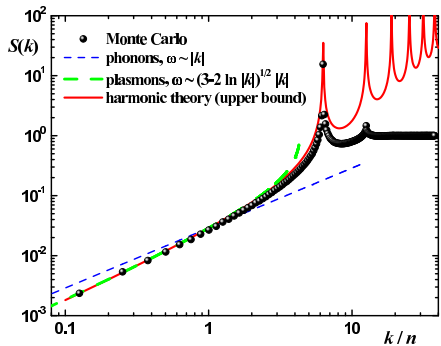}
\caption{(Color online) Static structure factor for $N=100$ particles in the Wigner crystal regime ($n a_0 = 0.01$). Symbols, DMC results; lines, Feynman relation $S_k = \hbar^2k^2/(2m\omega_k)$ for different dispersion relations: phonons and plasmons, Eq.~(\ref{plasmon}); harmonic theory $[\hbar\omega(k)/Ry]^2 = 4\zeta(3)-2Li_3(e^{ik/n})-2Li_3(e^{-ik/n})$ \cite{plasmon1D}.}
\label{Fig:Sk}
\end{center}
\end{figure}

The low-lying excitations in gases with short-range potentials are phonons with a linear dispersion relation $\omega_k = |k|c$, where $c$ is the speed of sound. This is the case of $\delta$-interacting Lieb-Liniger \cite{Lieb}, Calogero-Sutherland\cite{CSM}, and Tonks-Girardeau\cite{Girardeau60} gases, in which the long-range properties of the correlation functions are described by the Luttinger liquid theory\cite{Haldane}. The long-range nature of the Coulomb interaction potential leads to different low-lying excitations which instead are plasmons. Their dispersion relation can be found from the classical equations of motions in a Wigner crystal\cite{Schulz,plasmon1D}. The low-momentum dependence is nonanalytic and contains logarithmic terms
\begin{eqnarray}
\omega^2_k =
\frac{2e^2n}{m}\left|\ln\frac{k}{n}\right| k^2
+\frac{3e^2n}{m} k^2
+O(k^4)\ .
\label{plasmon}
\end{eqnarray}
When the excitation spectrum is exhausted by one branch of excitations, the excitation spectrum is related to the static structure factor $S_k=\langle\rho^\dagger_k\rho_k\rangle/N$ by the Feynman formula $S_k = \hbar^2k^2/(2m\omega_k)$. We perform a DMC calculation of the static structure factor and present the results in Fig.~\ref{Fig:Sk}. We observe that the low-energy charge excitations are well described by the plasmonic dispersion relation~(\ref{plasmon}). We note that although the logarithmic term in~(\ref{plasmon}) mathematically is the leading one, it alone is not enough to describe accurately the excitation spectrum and the use of the subsequent term is very important. In the opposite regime of large density the dispersion relation is similar to that of an ideal Fermi gas, and the low-lying excitations are phonons.

The long-range nature of the Coulomb potential causes no divergences in finite-size trapped systems. We consider the case of a harmonic oscillator and add the confinement term $V_{trap}(z) = m\omega^2z^2/2$ to Hamiltonian~(\ref{H}). It is convenient to use oscillator units of $\hbar\omega$ and oscillator length $a_{osc}=\sqrt{\hbar/m\omega}$ for energy and distances, respectively. In dimensionless units the Hamiltonian reads as
\begin{equation}
\hat{H}^{trap}_{dim}=
-\frac{1}{2}\sum_{i=1}^{N}\frac{\partial^2}{\partial z_i^2}
+\frac{1}{2}\sum_{i=1}^N z_i^2
+\sum_{i<j}^N\frac{q^2}{|z_i-z_j|}
\label{Htrapdim}
\end{equation}
and the system is characterized by the number of particles $N$ and the dimensionless charge $q=\sqrt{a_{osc}/a_{0}}$.

\begin{figure}
\begin{center}
\includegraphics[width=0.52\columnwidth, angle=-90]{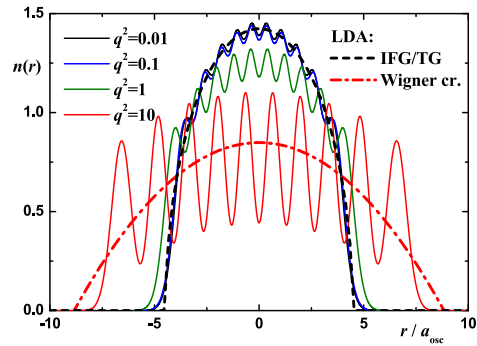}
\caption{(Color online) Density profile of $N=10$ trapped particles. Solid line, diffusion Monte Carlo result;
local density approximation with equation of state as in ideal Fermi gas (dashed line) and Wigner crystal (dash-dotted line). DMC density profiles are calculated using the technique of pure estimators \cite{Casulleras95}.}
\label{Fig3}
\end{center}
\end{figure}

For a small number of particles the density profile has a typical shell structure of a mesoscopic system as shown in Fig.~\ref{Fig3}. When the shell structure can be neglected, the system properties are well described within the local density approximation (LDA) \cite{dgps}. The chemical potential $\mu$ of the trapped system is then approximated by the sum of the external potential $V_{trap}(z)$ and the chemical potential $\mu_{hom}(n)$ of the homogeneous system, with the local density $n(z)$. The value of the chemical potential is fixed by the normalization condition $N=\int n(z)\;dz$.

In the limit of small charge $q\to 0$, the chemical potential of a homogeneous system as in Eq.~(\ref{ETG}) can be approximated by a kinetic energy of an ideal Fermi gas $\mu_{hom} = \pi^2\hbar^2n^2/2m$ and is independent of particle charges. The density profile is a semicircle of a Thomas-Fermi size $R=\sqrt{2N}a_{osc}$:
\begin{equation}
n(r) = \frac{1}{\pi}\sqrt{R^2-r^2}.
\label{LDATG}
\end{equation}
The chemical potential is $\mu = N\hbar\omega$ and the energy per particle is $E/N = N \hbar\omega/2$. Even if the LDA does not describe the shell structure, thus leaving out contributions of the order of $1/N$, the obtained LDA result for the energy turns out to be exact for a trapped ideal Fermi gas.

In the limit of large charge $q\to\infty$, the chemical potential of a homogeneous system has a linear dependence on the density $\mu_{hom} = 2e^2 n\ln N$, which is obtained by differentiating the Wigner crystal energy~(\ref{Ewigner}). The density profile has an inverted parabola shape, typical for gases with short-range interaction potentials\cite{dgps},
\begin{equation}
n(r) = \frac{1}{4q^2\ln N}(R^2-r^2)\ ,
\label{LDAMF}
\end{equation}
with Thomas-Fermi radius $R = (3N\ln N q^2)^{1/3}$. The chemical potential is $\mu = 1/2\;(3q^2 N\ln N)^{2/3}$ and the energy per particle $E/N = 3/10\;(3q^2 N\ln N)^{2/3}$.

We adapt the guiding wave function~(\ref{wf}) of a homogeneous system by multiplying it by one-body terms $f_1(z_i) = \exp(-\alpha z_i^2)$ with the free parameter $\alpha$ optimized by minimizing the variational energy. The two-body terms are chosen as $f_2(z) = \sqrt{z}I_{1}(2\sqrt{q z})$.

\begin{figure}
\begin{center}
\includegraphics[width=0.52\columnwidth, angle=-90]{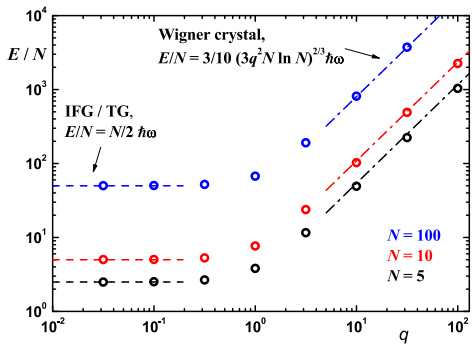}
\caption{(Color online) Ground-state energy of $N= 100; 10; 5$ (from top to bottom) charged particles in a harmonic trap. Circles, DMC results; dashed line, energy if IFG/TG gas; dash-dotted line, LDA result for Wigner crystal.}
\label{Fig4}
\end{center}
\end{figure}

Figure~\ref{Fig4} shows the energy of $N$ charged particles in a harmonic trap. For small charges the energy per particle is well described by that of an ideal Fermi gas and increases linearly with the number of particles $N$. Each new particle occupies the next $l=N-1$ level of the harmonic oscillator and the energy increases by $E_l = (l+1/2)\hbar\omega$. For large charges the local density approximation result with Wigner crystal equation of state applies. In this regime the energy per particle increases as $(N\ln N)^{3/2}$. In between the two limits there is a smooth crossover.

The density profile for a system of $N=10$ is presented in Fig.~\ref{Fig3}. For small charges the LDA predicts a semicircular shape of a radius which depends only on the number of particles and is independent of the value of the charge. For that reason different density profiles with $q^2\lesssim 0.1$ almost coincide. The regime of large charges is a trap analogue of a Wigner crystal. The particles are highly localized and form well defined shells, while the overall shape is an inverted parabola of a width which depends on the value of the charge. Differently from a homogeneous Wigner crystal, here the distance $a$ between lattice sites is no longer constant and depends on the distance $z$ from the center of the trap as $a(z) \approx 1/n(z)$. The LDA description is known to be imprecise at the borders of the system\cite{dgps} as it predicts cusps in the density, while the true density profile is smooth as can be seen in Fig.~\ref{Fig3}. Furthermore, the large-density Wigner crystal equation of state is not applicable at the edges where the density is small, which worsens the LDA description of a trapped Wigner crystal at the edges. The square of the size of the system is related to the potential energy of the trap and has a similar behavior to the total energy, see Fig.~\ref{Fig4}.


To conclude, we have studied properties of a single component Coulomb $1/r$ gas in a one-dimensional geometry. The short-range divergence of the interaction potential permits us to exploit the Fermi-Bose mapping for the ground state wave function and to solve the Fermi sign problem in the considered system as (i) the mapping allows exact evaluation of the ground state properties, (ii) the numerical complexity of a simulation in a fermionic system is the same as in a bosonic system. This permits us to carry out {\it ab initio} simulation of many-body systems with as many as 1000 fermions. Due to the mapping, the bosonic and fermionic Coulomb systems have the same energy and local properties (density profile, pair distribution functions, etc.).

We use the diffusion Monte Carlo method to do an exact calculation of the ground state energy of a system in a box with periodic boundary conditions and in a harmonic trap. In the regime of large density a homogeneous system is well described by an ideal Fermi gas for fermions or Tonks-Girardeau gas for bosons. There is a smooth crossover transition to the Wigner crystal regime as the density is lowered. We show that the low-lying excitations are plasmons as manifested in the low-momentum behavior of the static structure factor.

The energy and the density profile in trapped systems are obtained within the local density approximation and are confronted with the exact results of the diffusion Monte Carlo calculation. In the ideal Fermi gas/Tonks-Girardeau limit the system properties are defined by the number of particles and are independent of the charge. The limit of large charge is a trapped analogue of a Wigner crystal where localized shells play a similar role to crystal lattice sites. The amplitude of the oscillations in the density profile experiences a drastic change in the shape varying from a semicircle (IFG/TG limit) to an inverted parabola (Wigner crystal limit).

G.E.A. acknowledges support from the Spanish MEC through the Ramon y Cajal fellowship program.

\end{document}